\documentclass[11pt,twoside]{article}
\usepackage[pdftex]{graphicx}
\usepackage{amsmath}
\usepackage{amssymb}
\usepackage{cite}

\begin{document}
\newcommand{\pst}{\hspace*{1.5em}}

\newcommand{\rigmark}{\em Journal of Russian Laser Research}
\newcommand{\lemark}{\em Volume 30, Number 5, 2009}

\newcommand{\be}{\begin{equation}}
\newcommand{\ee}{\end{equation}}
\newcommand{\bm}{\boldmath}
\newcommand{\ds}{\displaystyle}
\newcommand{\bea}{\begin{eqnarray}}
\newcommand{\eea}{\end{eqnarray}}
\newcommand{\ba}{\begin{array}}
\newcommand{\ea}{\end{array}}
\newcommand{\arcsinh}{\mathop{\rm arcsinh}\nolimits}
\newcommand{\arctanh}{\mathop{\rm arctanh}\nolimits}
\newcommand{\bc}{\begin{center}}
\newcommand{\ec}{\end{center}}

\thispagestyle{plain}

\label{sh}


\begin{center} {\Large \bf
\begin{tabular}{c}
INEQUALITIES FOR NONNEGATIVE NUMBERS\\[-1mm]
AND INFORMATION PROPERTIES\\[-1mm]
OF QUDIT TOMOGRAMS
\end{tabular}}
\end{center}

\bigskip

\begin{center} {\bf
Margarita A. Man'ko$^{\ast}$ and Vladimir I. Man'ko}

\medskip

{\it
P. N. Lebedev Physical Institute, Russian Academy of Sciences\\
Leninskii Prospect 53, Moscow 119991, Russia
}
\smallskip

$^*$Corresponding author e-mail:~~~mmanko\,@\,sci.lebedev.ru\\
\end{center}

\begin{abstract}\noindent
We discuss some inequalities for $N$ nonnegative numbers. We use these
inequalities to obtain known inequalities for probability distributions and
new entropic and information inequalities for quantum tomograms of qudit
states. The inequalities characterize the degree of quantum correlations in
addition to noncontextuality and quantum discord. We use the subadditivity
and strong subadditivity conditions for qudit tomographic-probability
distributions depending on the unitary-group parameters in order to derive
new inequalities for Shannon, R\'enyi, and Tsallis entropies of spin
states.
\end{abstract}

\medskip


\section{Introduction}
\pst There exist quantum phenomena related to the presence of quantum
correlations. The quantum correlations are responsible for
entanglement~\cite{Schrod}, violation of the Bell
inequalities~\cite{Bell,Bella,Horn}, noncontextuality (see, for example,
\cite{Shumov,Lapki}), and discord~\cite{Zurek}. In some cases, the
existence of quantum correlations can be expressed in terms of the
tomographic-probability distributions (spin
tomograms)~\cite{OlgaJETP,DodPLA,SudarPLA,Ibort} and their specific
properties. These properties were discussed for Shannon
entropy~\cite{Shannon} and $q$-entropy~\cite{Renyi,Tsallis} and information
associated with the tomographic-probability distributions in
\cite{ActaHung06,EJPB06,VinoSem,JPCS07FeynFest,TMP07,AIP07MA,VovaJRLR08,
TMP09,RitaJRLR09,RitaPSenin,RitaPScomt,RitaTMF11,RitaJRLR11,RitaFP,RitaPS12}.

The idea of our approach is to consider three different but closely
connected objects. The probability distributions are determined by a set of
nonnegative numbers. In view of this fact, our first object is the set of
nonnegative numbers not related to any applications. It is a purely
mathematical object with specific properties. These properties can be
studied considering some functions on the set of nonnegative numbers. The
functions can satisfy some inequalities that are generic inequalities
characterizing both the set of nonnegative numbers and the functions. The
second object is the standard probability distributions that are identified
with the set of nonnegative numbers with an additional interpretation that
these numbers are the probability of some observable measurements, and the
observables themselves are associated with other numbers, which code the
results of the measurements. The third object to be considered is the
nonnegative functions defined, for example, on unitary matrices. For each
unitary matrix (or a point on the sphere), one has the discussed set of
nonnegative numbers. Also the nonnegative functions can be associated with
the probability distributions considered as the probability distributions
depending on extra parameters like the unitary matrices or point on the
sphere. This means that the parameter-dependent probability distributions
are the nonnegative functions, and for each parameter there exists another
set of numbers that code the outcome of experiments where some observables
are measured.

We try to consider entropic and information inequalities analyzing what
system properties are connected with only mathematical properties of the
sets of nonnegative numbers and what properties are associated with an
extra information contained in the probability distributions of measurable
variables and the results of experiments for the cases where the dependence
on some parameters like unitary matrices or coordinates of a point on the
sphere play a role.

The aim of this paper is to connect the entropic and information
inequalities (uncertainty relations) with some general properties of a set
of $N$ positive numbers and properties of unitary matrices.

This paper is organized as follows.

In Sec.~2, we discuss the properties of nonnegative numbers and some
inequalities for these numbers and consider the interpretation of the
nonnegative numbers in terms of the probability distributions. In Sec.~3,
we apply the obtained results to the tomographic-probability distributions
of quantum systems. In Sec.~4, we study a qudit system and consider the
Shannon and $q$-entropies in Sec.~5. In Sec.~6, we review known entropic
inequalities and obtain new information inequalities in Sec.~7. In Sec.~8,
we study the probability properties, in view of the vector and matrix
properties. In Sec.~9, we discuss the influence of permutations of
nonnegative numbers on the properties of entropies and consider the
relation between the strong subadditivity condition and matrices in
Sec.~10. Our conclusions are presented in Sec~11.

\section{Nonnegative Numbers}
\pst We consider a set of $N$ nonnegative numbers $P_1,P_2,\ldots,P_N$. Let
these numbers satisfy the additional normalization condition
$\sum_{k=1}^NP_k=1$. There are different functions $f(P_1,P_2,\ldots,P_N)$,
which have the index permutation symmetry, i.e.,
$f(P_1,P_2,\ldots,P_N)=f(\hat{\cal P}_{P_1},\hat{\cal P}_{P_2}, \ldots,
\hat{\cal P}_{P_N})$, where $\hat{\cal P}_{P_j}$ means the result of a
permutation-operator action on the $j$th nonnegative number. The
Shannon~\cite{Shannon} and R\'enyi~\cite{Renyi} entropies have such a
symmetry in the case where the nonnegative numbers $P_j$ are associated
with the probability distributions describing the results of measurements
in different nondeterministic processes. Independently of the
interpretation in terms of the probability distributions, it is worth
pointing out that the set of positive numbers can be characterized by some
inequalities for functions that can be considered in the applications as
entropies, information, etc.

The permutation of the nonnegative numbers can be visualized if the numbers
are organized in a vector $\vec P$. Then permutation is described by a
stochastic matrix acting on this vector. Thus, there exist $N!$ different
permutation $N$$\times$$N$-matrices. The permutation matrices are unitary
real matrices with matrix elements equal either to zero or to unity. We
discuss also the action of the permutations onto an nonnegative Hermitian
$N$$\times$$N$-matrix considered as a complex $N^2$-vector with components
constructed of rows of the matrix~\cite{SudarJRLR}. Then the permutation
$N^2$$\times$$N^2$-matrix acting on the vector is the direct product of two
unitary permutation $N$$\times$$N$-matrices. The matrix realizes a specific
positive map of the Hermitian matrix. The map does not change the matrix
eigenvalues but yields the permutation of the matrix eigenvectors.

\section{Joint Probability Properties}
\pst In this section, we discuss the probability distributions on the
example of a bipartite system consisting of two subsystems.

It is known that, if one has the joint probability distribution
$w(m_1,m_2)\geq 0$ of two discrete random variables $m_1$ and $m_2$ related
to a system containing two subsystems 1 and 2, there exist two marginals
\begin{equation}\label{JP1}
P_1(m_1)=\sum_{m_2}w(m_1,m_2),\qquad P_2(m_2)=\sum_{m_1}w(m_1,m_2),
\end{equation}
which are associated with Shannon entropies
\begin{equation}\label{JP2}
H(k)=-\sum_{m_k}P_k(m_k)\ln P_k(m_k)\geq 0,\qquad k=1,2.
\end{equation}
The entropy of the system $H(1,2)$ reads
\begin{equation}\label{JP3}
H(1,2)=-\sum_{m_1,m_2}w(m_1,m_2)\ln w(m_1,m_2)\geq 0.
\end{equation}
There exists the inequality called the subadditivity condition
\begin{equation}\label{JP4}
H(1)+H(2)\geq H(1,2),
\end{equation}
and the Shannon mutual information is defined as the difference
\begin{equation}\label{JP5}
I=H(1)+H(2)-H(1,2)\geq 0.
\end{equation}
There exist other probability distributions determined by the initial
distribution $w(m_1,m_2)$. For example, two conditional probability
distributions ${\cal P}_1(m_1\mid m_2)$ and ${\cal P}_2(m_2\mid m_1)$ are
defined as
\begin{equation}\label{JP6}
{\cal P}_1(m_1\mid m_2)= \frac{P_1(m_1,m_2)}{P_2(m_2)}\,,\qquad {\cal
P}_2(m_2\mid m_1)=\frac{P_1(m_1,m_2)}{P_1(m_1)}\,.
\end{equation}
The meaning of the conditional probability distribution follows from the
obvious statement, which is the essence of the Bayesian
formula~(\ref{JP6}), namely, the joint probability $P(m_1,m_2)$ to obtain
the values of two random variables $m_1$ and $m_2$ (measurable
simultaneously) is equal to the product of the probability $P_2(m_2)$ to
obtain the variable $m_2$ and the probability ${\cal P}_1(m_1\mid m_2)$ to
obtain the value $m_1$ of the first random variable under the condition
that the value of variable $m_2$ is known.

The conditional probability distributions determine the Shannon entropies
\begin{eqnarray}\label{JP7}
H_1(1\mid m_2)=-\sum_{m_1}{\cal P}_1(m_1\mid m_2)\ln {\cal P}_1(m_1\mid
m_2),\nonumber\\[-2mm]
\label{JP7}\\[-2mm]
H_2(2\mid m_1)=-\sum_{m_2}{\cal P}_2(m_2\mid m_1)\ln {\cal P}_2(m_2\mid
m_1).\nonumber
\end{eqnarray}
One can calculate average entropies
\begin{eqnarray}
H(1\mid 2)=\overline {H_1}=\sum_{m_2}P_2(m_2)H_1(1\mid m_2), \nonumber\\[-2mm]
\label{JP8}\\[-2mm]
 H(2\mid
1)=\overline {H_2}=\sum_{m_1}P_1(m_1)H_2(2\mid m_1).\nonumber
\end{eqnarray}
One can check that the following equalities are valid:
\begin{equation}\label{JP9}
\overline {H_1}=H(1,2)-H(2)=H(1\mid 2),\qquad \overline
{H_2}=H(1,2)-H(1)=H(2\mid 1).
\end{equation}
Also the mutual information can be expressed as the difference
\begin{equation}\label{JP10}
I=-\overline {H_1}+H(1)=-\overline {H_2}+ H(2).
\end{equation}
The nonnegativity of the mutual information means that
\begin{equation}\label{JP11}
H(1)\geq\overline {H_1},\qquad H(2)\geq \overline {H_2}.
\end{equation}

\subsection{Example of the Classical Coins}
\pst We illustrate the discussed notions on the example of two classical
coins.

Let the first and second coins have the outcomes of the experiment (up and
down) labeled by $\pm 1$. This means that one has the probability
distribution $w(m_1,m_2)$ determined by four nonnegative numbers
$$w(+1,+1)=a,\qquad w(+1,-1)=b,\qquad w(-1,+1)=c,\qquad w(-1,-1)=d.
$$
The normalization of the probability distribution $w(m_1,m_2)$ means that
$a+b+c+d=1.$ One can consider the probability distribution $w(m_1,m_2)$ as
a probability column vector $\vec w$ with four components. The marginals
$P_1(m_1)$ and $P_2(m_2)$ are the probability distributions
$$P_1(+1)=a+b,\qquad P_1(-1)=c+d,\qquad P_2(+1)=a+c,\qquad P_2(-1)=b+d.$$
These marginals can be considered either as the probability column vectors
$\vec P_1$ and $\vec P_2$,
$$\vec P_{1}=\left(\begin{array}{c}
            a+b\\
            c+d \\
          \end{array}
        \right),\qquad\vec P_{2}=\left(\begin{array}{c}
            a+c\\
            b+d \\
          \end{array}
        \right),$$
or as the probability column 4-vectors obtained as the qubit
portrait~\cite{VovaJRLR,LupoJPA,FilipJRLR} of the initial vector $\vec w$,
using stochastic matrices
$$
M_{1}=\left(\begin{array}{cccc}
            1&1&0&0\\
0&0&1&1\\0&0&0&0\\0&0&0&0\\
          \end{array}
        \right),\qquad
M_{2}=\left(\begin{array}{cccc}
            1&0&1&0\\
0&1&0&1\\0&0&0&0\\0&0&0&0\\
          \end{array}
        \right). $$
Thus, one has $\vec\Pi_1=M_1\vec w$ and $\vec\Pi_2=M_2\vec w$, where the
column vectors $\vec\Pi_1$ and $\vec\Pi_2$ read $$\vec
\Pi_{1}=\left(\begin{array}{c}
            \vec P_1\\
            \vec 0 \\
          \end{array}
        \right), \qquad\vec \Pi_{2}=\left(\begin{array}{c}
\vec P_2\\
            \vec 0 \\
          \end{array}
        \right), \qquad\vec 0=\left(\begin{array}{c}
            0\\
            0 \\
          \end{array}
        \right).$$

\subsection{The Probability Vector as a Rectangular Matrix}
\pst Now we formulate the general matrix rule for constructing the
marginals for the probability $N$-vector $\vec P$ given as a column with
$N=mn$ components, i.e., $$\vec P=\big(P_{11},P_{12},\ldots
P_{1n}P_{21},P_{22},\ldots,P_{2n}\ldots,P_{m1},P_{m2},\ldots,P_{mn}\big).$$
First we represent $\vec P$ in the form of a rectangular matrix
\begin{equation}\label{JP12}
P_{kj}=\left(\begin{array}{cccc}
            P_{11}& P_{12}& \ldots & P_{11}\\
 P_{21}& P_{22}& \ldots & P_{1n}\\
\ldots &\ldots &\ldots &\ldots \\
 P_{21}& P_{22}& \ldots& P_{2n}\\
 P_{m1}& P_{m1}& \ldots& P_{mn}\\
          \end{array}
        \right).
\end{equation}
Then the two marginals are represented by the probability $m$-vector $\vec
P_1$ and $n$-vector $\vec P_2$. The components of the $m$-vector $\vec P_1$
are obtained by the sum of the matrix elements in the $k$th rows,
\begin{equation}\label{JP13}
\big(\vec P_{1}\big)_k=\sum_{k=1}^nP_{ks},\qquad k=1,2,\ldots,m,
\end{equation}
and the components of the $n$-vector $\vec P_2$ are obtained by the sum of
the matrix elements in the $j$th columns,
\begin{equation}\label{JP14}
\big(\vec P_{2}\big)_j=\sum_{l=1}^mP_{lj},\qquad j=1,2,\ldots,n.
\end{equation}
The subadditivity condition (\ref{JP4}) can be formulated as an inequality
for the matrix $P_{kj}$~(\ref{JP12}) following the statement: Given a
rectangular matrix with nonnegative matrix elements $P_{kj}$ such that
$\sum_k\sum_jP_{kj}=1$, one has
\begin{equation}\label{JP15}
\sum_{j=1}^n\left[\left(\sum_{l=1}^mP_{lj}\right)\ln\sum_{l'=1}^mP_{l'j}
\right]+\sum_{k=1}^m\left[\left(\sum_{s=1}^nP_{ks}\right)\ln\sum_{s'=1}^nP_{ks'}
\right]\leq\sum_j^n\sum_k^mP_{kj}\ln P_{kj}.
\end{equation}
Entropy~(\ref{JP3}) reads $$H(1,2)=-a\ln a-b\ln b-c\ln c- d\ln d\equiv
-\vec w\ln\vec w,$$ and the entropy associated with the marginals are
\begin{eqnarray}\label{JP16}
H(1)=-(a+b)\ln(a+b)-(c+d)\ln(c+d)=-\vec\Pi_1\ln\vec\Pi_1=-(M_1\vec
w)\ln(M_1\vec w),\nonumber\\[-2mm]
\\[-2mm] H(2)=-(a+c)\ln(a+c)-(b+d)\ln(b+d) =-\vec\Pi_2\ln\vec\Pi_2=-(M_2\vec
w)\ln(M_2\vec w).\nonumber
\end{eqnarray}
We use here the notation $\ln\vec A=\vec A_{\ln}$, which means the
$N$-vector with components $(\ln\vec A)_k\equiv\ln A_k$. Also we use for
any function $f(x)$ the notation for a vector $\vec A_f\equiv f(\vec A)$
with components $(\vec A_f)_k\equiv f(A_k)$, $~k=1,2,\ldots,N$. Precisely
in our case, for the Shannon entropy the function $f(x)=\ln x$. The scalar
product of real vectors $\vec A_f\vec B_\varphi$ is defined as
$$ f(\vec A)\varphi(\vec B)\equiv\vec A_f\vec
B_\varphi=\sum_{k=1}^nf(A_k)\varphi(B_k). $$

\subsection{Conditional Probability Distribution for Two Coins}
\pst The mutual information is given by the expression
\begin{eqnarray}\label{JP17}
I&=&\vec w\ln\vec w-(M_1\vec w)\ln(M_1\vec w)-(M_2\vec w)\ln(M_2\vec
w)\nonumber\\
&=&a\ln a+b\ln b+c\ln c+d\ln d-(a+b)\ln(a+b)-(c+d)\ln(c+d)\nonumber\\
&& -(a+c)\ln(a+c)-(b+d)\ln(b+d).
\end{eqnarray}
We make a general statement that follows from inequality~(\ref{JP15}).

Given a probability vector $\vec P$, i.e., $N$ nonnegative numbers
$P_\alpha$, $~\sum_{\alpha=1}^NP_\alpha=1$, we distribute these numbers in
any way in a rectangular matrix $P_{kj}$ with the number of matrix elements
larger than $N$ and put the number zero for matrix elements in empty
positions. Then we have inequality~(\ref{JP15}).

The conditional probability distributions ${\cal P}_1(m_1\mid m_2)$ read
\begin{eqnarray}\label{JP18}
{\cal P}_1(+1\mid+1)=\frac{a}{a+c}\,,\qquad {\cal
P}_1(-1\mid+1)=\frac{c}{a+c}\,,\nonumber\\[-2mm]
\\[-2mm] {\cal P}_1(+1\mid-1)=\frac{b}{b+d}\,,\qquad {\cal
P}_1(-1\mid-1)=\frac{d}{b+d}\,,\nonumber
\end{eqnarray}
and the two entropies are
\begin{eqnarray}\label{JP19}
H(+1\mid+1)=\left(-\frac{a}{a+c}\ln\frac{a}{a+c}-\frac{c}{a+c}\ln\frac{c}{a+c}\right),
\nonumber\\[-2mm]
\\[-2mm]
H(+1\mid-1)=\left(-\frac{b}{b+d}\ln\frac{b}{b+d}-\frac{d}{b+d}\ln\frac{d}{b+d}\right).\nonumber
\end{eqnarray}

We define the entropy ${\overline H}(1)=H(1\mid -1)$ as
\begin{eqnarray}\label{JP20}
{\overline H}(1)&=&H(1\mid +1)P_2(+1)+H(1\mid -1)P_2(-1)\nonumber\\
&=&
\left(-\frac{a}{a+c}\ln\frac{a}{a+c}-\frac{c}{a+c}\ln\frac{c}{a+c}\right)(a+c)
\nonumber\\
&&+\left(-\frac{b}{b+d}\ln\frac{b}{b+d}-\frac{d}{b+d}\ln\frac{d}{b+d}\right)(b+d)\nonumber\\
&=&-{a}\ln\frac{a}{a+c}-{c}\ln\frac{c}{a+c}-{b}\ln\frac{b}{b+d}-{d}\ln\frac{d}{b+d}\nonumber\\
&=&H(1,2)-H(2).
\end{eqnarray}
Analogously,
\begin{equation}\label{JP21}
{\overline H}(2)=H(2\mid 1)=H(1,2)-H(1),
\end{equation}
and we obtain the following rule.

We consider a probability vector $\vec P$ with $N$ components
$P_1,P_2,\ldots P_N$ and construct the portrait of this probability vector,
which is a new probability vector $\vec \Pi$ given by the action of the
fiducial stochastic matrix $M_f$, such that
$$\Pi_1=\big(P_1+P_2+\cdots+P_{j_1}\big),
\Pi_2=\big(P_{j_1+1}+P_{J_1+2}+\cdots+P_{j_2}\big),\ldots,
\Pi_s=\big(P_{j_{s-1}+1}+P_{j_{s-1}+2}+\cdots+P_N\big),$$ and all the other
vector components are zeros. Such a map provides the Shannon entropy of the
portrait probability distributions
$$ H_\Pi=-\sum_{k=1}\Pi_k\ln\Pi_k.$$
Also there are $s_N$ conditional probability distributions created by the
map $M$, namely,
\begin{eqnarray}\label{JP22}
{\cal P}(1\mid 1)=P_1/\Pi_1,~{\cal P}(2\mid 1)=P_2/\Pi_1,~\ldots,{\cal
P}(j_1\mid 1)=P_{j_1}/\Pi_1,\nonumber\\
{\cal P}(1\mid 2)=P_{j_1+1}/\Pi_2,~{\cal P}(2\mid
2)=P_{j_1+2}/\Pi_2,~\ldots,{\cal
P}(j_2\mid 2)=P_{j_2}/\Pi_2,\\
\ldots\ldots\ldots\ldots\ldots\ldots\ldots
\ldots\ldots\ldots\ldots\ldots\ldots\ldots\ldots
\ldots\ldots\ldots\ldots\ldots\ldots\nonumber\\
 {\cal P}(1\mid s)=P_{j_s+1}/\Pi_s,~{\cal P}(2\mid
s)=P_{j_s+2}/\Pi_s,~\ldots,{\cal P}(j_s\mid s)=P_{N}/\Pi_s,\nonumber
\end{eqnarray}
where $j_1+j_2+\cdots +j_s=N.$

The entropies defined as analogs of the entropies associated with
conditional probability distributions read $$H(k)=-\sum_l{\cal P}(l\mid
k)\ln{\cal P}(l\mid k).$$ Then the average entropy $\overline H$ is
expressed as $$\overline H=H-H_\Pi=-\vec P\ln\vec P+\vec\Pi\ln\vec\Pi.$$
The conditional probability distribution means the probability distribution
to have the outcome of the event if it is known that the event belongs to
the $k$th group given by the $k$th row of the portrait matrix $M_f$ of the
stochastic matrix.

\section{Spin Tomograms (Qubit and Qudit Tomograms)}
\pst Given an $N$-dimensional space of states of spin system. One can
interpret this space either as the state space for one particle with spin
$j=(N-1)/2$ (qudit) or, in the case of the product representation of the
number $N=n_1n_2\cdots n_M$, as the space of multipartite spin system
(multipartite qudit system) with $j_1=(n_1-1)/2$,
$j_2=(n_2-1)/2,\ldots,j_M=(n_M-1)/2$.

The $N$$\times$$N$ density matrix $\rho$ of the quantum state can be
represented by the unitary tomogram of the spin state~\cite{SudarPLA}. In
the case of the spin state with $j=(N-1)/2$, the tomogram is defined by the
relation
\begin{equation}\label{1}
w(m,u)=\langle m\mid u^\dagger\rho u\mid m\rangle,
\end{equation}
where $\rho$ is the density matrix, $u$ is the $N$$\times$$N$ unitary
matrix, and semi-integers $m=-j,-j+1,\ldots,j$ are values of the spin
projection on the $z$ axis. Tomogram~(\ref{1}) is the nonnegative
probability-distribution function of the random spin-projection variable
satisfying the normalization condition $\sum_{m=-j}^jw(m,u)=1$ and the
equality $\int w(m,u)\,du=1$, where $du$ is the Haar measure on the unitary
group with the normalization $\int du=1.$ An important property of tomogram
$w(m,u)$ is that its connection with the density matrix $\rho$ reads
$\rho\leftrightarrow w(m,u).$ This means that the quantum state is given if
the tomogram is known~\cite{OlgaJETP,DodPLA}.

\section{Shannon and R\'{e}nyi Tomographic Entropies}
\pst Following standard definitions of the probability theory, one can
introduce Shannon~\cite{Shannon} tomographic entropy~\cite{Olga2} and
R\'{e}nyi~\cite{Renyi} tomographic entropy~\cite{Rui-JRLR}.

The Shannon tomographic entropy is the function on the unitary group
\begin{equation}\label{12}
H_{u}=-\sum_{m=-j}^jw(m,u)\ln w(m,u).
\end{equation}
The R\'{e}nyi tomographic entropy is also the function on the unitary group
and it depends on an extra parameter $q$
\begin{equation}\label{13}
R_{u}^{(q)}=\frac{1}{1-q}\ln
\left(\sum_{m=-j}^j\big(w\left(m,u\right)\big)^q \right). \end{equation}
The Tsallis tomographic entropy is determined as
\begin{equation}\label{Ts}
T_{u}(q)=\frac{1}{1-q}\left(\sum_{m=-j}^j\big(w\left(m,u\right)\big)^q-1\right).
\end{equation}

For two spin tomograms $w_1(m,u)$ and $w_2(m,u)$, we define the relative
tomographic $q$-entropy
\begin{equation}\label{14}
H_{q}\left(w_{1}(u)|w_{2}(u)\right)=-\sum_{m=-j}^jw_1(m,u)\ln _{q}
\frac{w_{2}(m,u)}{w_{1}(m,u)},
\end{equation} with
$$\ln_qx=\frac{x^{1-q}-1}{1-q}\,,\qquad x>0,\qquad q>0,\qquad
\ln_{q\rightarrow 1}x=\ln x.$$ The relative tomographic $q$-entropy is a
nonnegative function for any admissible deformation parameter $q$. For
$q\to 1$, $R_u\to H_u$ and the relative tomographic $q$-entropy becomes the
relative entropy associated to the two tomographic-probability
distributions
\begin{equation}\label{16}
H\big(w_{1}(u)|w_{2}(u)\big)=-\sum_{m=-j}^jw_1(m,u)\ln\frac{w_{2}(m,u)}{w_{1}(m,u)}.
\end{equation}
As was shown in \cite{Rui-JRLR}, the minimum over the unitary group of the
R\'{e}nyi tomographic entropy~(\ref{13}) is equal to the quantum R\'{e}nyi
tomographic entropy
\begin{equation}\label{17}
\min\,R_{u}=\frac{1}{1-q}\ln \mbox{Tr}\,\rho^q.
\end{equation}
The minimum over the unitary group of the Shannon tomographic
entropy~(\ref{12}) is equal to the von Neumann
entropy~\cite{Olga2,Rui-JRLR}, i.e.,
\begin{equation}\label{18}
\min\,H_{u}=-\mbox{Tr}\,\rho\,\ln\rho.
\end{equation}
One has for $\min\,R_{u}$ the corresponding quantum Tsallis entropy
\begin{equation}\label{Ts}
\frac{1}{1-q}\left(\mbox{Tr}\,\rho^q-1\right)=
\frac{1}{1-q}\left\{\exp[\min R_u(1-q)]-1\right\}.
\end{equation}

\section{Known Inequalities for Bipartite and Tripartite Systems}
\pst The tomographic entropies satisfy some known inequalities found in
\cite{Rui-JRLR}.

For example, if the spin system is bipartite, i.e., one has spins $j_1$ and
$j_2$, the basis in the tensor-product space reads $\mid m_1m_2\rangle=\mid
m_1\rangle\mid m_2\rangle.$ In this case, the tomogram is the
joint-probability distribution of two random spin projections
$m_1=-j_1,-j_1+1,\ldots,j_1$ and $m_2=-j_2,-j_2+1,\ldots,j_2$ depending on
the $(2j_1+1)(2j_2+1)$$\times$$(2j_1+1)(2j_2+1)$ unitary matrix $u$. The
tomogram reads
\begin{equation}\label{20}
w(m_1,m_2,u)=\langle m_1m_2\mid u^\dagger\rho(1,2)u\mid m_1m_2\rangle,
\end{equation}
where $\rho(1,2)$ is the density matrix of the bipartite-system state with
matrix elements
\begin{equation}\label{21}
\rho(1,2)_{m_1m_2,m'_1m'_2}=\langle m_1m_2\mid\rho(1,2)\mid
m'_1m'_2\rangle.
\end{equation}
For this tomogram, one can introduce the Shannon entropy $H_{12}(u)$ as
\begin{equation}\label{22}
H_{12}(u)=-\sum_{m_1=-j_1}^{j_1}\sum_{m_2=-j_2}^{j_2} w(m_1,m_2,u)\ln
w(m_1,m_2,u).
\end{equation}
The Shannon entropy $H_{12}(u)$ satisfies the subadditivity condition for
all elements of the unitary group
\begin{equation}\label{23}
H_{12}(u)\leq H_{1}(u)+H_{2}(u),
\end{equation}
where $H_{1}(u)$ and $H_{2}(u)$ are Shannon entropies associated with
subsystem tomograms
\begin{equation}\label{24}
w_1(m_1,u)=\sum_{m_2=-j_2}^{j_2}w(m_1,m_2,u),\qquad
w_2(m_2,u)=\sum_{m_1=-j_1}^{j_1}w(m_1,m_2,u)
\end{equation}
as follows:
\begin{equation}\label{26}
H_k(u)=-\sum_{m_k=-j_k}^{j_k}w_k(m_k,u)\,\ln w_k(m_k,u), \qquad k=1,2.
\end{equation}
From this inequality, in view of the relation between the von Neumann and
tomographic entropies, follows the known inequality~\cite{Rui-JRLR},
namely, the subadditivity condition for corresponding von Neumann entropy
for the bipartite system
\begin{equation}\label{27}
S_{12}\leq S_{1}+S_{2},
\end{equation}
where
\begin{equation}\label{28}
S_{k}=-\mbox{Tr}\,\rho_k\,\ln\rho_k, \quad k=1,2\qquad
\rho_{1}=-\mbox{Tr}_2\,\rho(1,2), \qquad \rho_{2}=-\mbox{Tr}_1\, \rho(1,2).
\end{equation}

For tripartite spin system with spins $j_1$, $j_2$, and $j_3$ and the
density matrix $\rho(1,2,3)$, the spin tomogram reads
\begin{equation}\label{30}
w(m_{1},m_{2},m_3,u)=\langle m_{1}m_{2}m_3\mid u^{\dagger }
\rho(1,2,3)u\mid m_{1}m_{2}m_3\rangle.
\end{equation}
One associates the Shannon entropy $H_{123}(u)$ with this tomogram. This
entropy satisfies the inequality, which is the strong subadditivity
condition on the unitary group. It reads~\cite{Rui-JRLR}
\begin{equation}\label{31}
H_{123}(u)+H_2(u)\leq H_{12}(u)+H_{23}(u),
\end{equation}
where
\begin{equation}\label{32}
H_{123}(u)=-\sum_{m_1=-j_1}^{j_1}\sum_{m_2=-j_2}^{j_2}\sum_{m_3=-j_3}^{j_3}
w(m_1,m_2,m_3,u)\ln w(m_1,m_2,m_3,u)
\end{equation} and entropies $H_{12}(u)$, $H_{23}(u)$, and $H_{2}(u)$ are defined by
means of projected tomograms
\begin{eqnarray*}
&&w_{12}(m_1,m_2,u)=\sum_{m_3=-j_3}^{j_3}w(m_1,m_2,m_3,u), \qquad
w_{23}(m_2,m_3,u)=\sum_{m_1=-j_1}^{j_1}w(m_1,m_2,m_3,u),\\
&&\qquad \qquad \qquad \qquad \qquad \qquad
w_{2}(m_2,u)=\sum_{m_1=-j_1}^{j_1}w_{12}(m_1,m_2,u).
\end{eqnarray*}
Our new inequality~(\ref{31}) is compatible with the known strong
subadditivity condition for the von Neumann entropy presented in
\cite{RusLieb,Rus}
\begin{equation}\label{36}
S_{123}+S_2\leq S_{12}+S_{23},\end{equation} where
$S_{123}=-\mbox{Tr}\,\rho_{123}\,\ln\rho_{123}$,
and other entropies are von Neumann entropies for reduced density matrices
$\rho(1,2)=\mbox{Tr}_3\rho(1,2,3)$ and $\rho(2,3)=\mbox{Tr}_1\rho(1,2,3)$.

Inequalities (\ref{23}) and (\ref{31}) are new inequalities for composite
quantum finite-dimensional systems obtained in \cite{Rui-JRLR}.

\section{Quantum Correlations and New Local-Transform Dependent \newline Information
Inequalities}
\pst In view of (\ref{23}), the Shannon tomographic
information is defined as
\begin{equation}\label{A}
I(u)= H_{1}(u)+H_{2}(u)-H_{12}(u),
\end{equation}
and in view of (\ref{27}), the quantum information is defined as
\begin{equation}\label{B}
I_q=S_{1}+S_{2}-S_{12}.
\end{equation}
If we consider equality~(\ref{A}) for the unitary matrix $u=u_{10}\otimes
u_{20}$, corresponding to local unitary transforms $u_{10}$ and $u_{20}$
for which $H_1(u_{10})=S_1$ and $H_2(u_{20})=S_2,$ i.e., the unitary
matrices $u_{10}$ and $u_{20}$ are acting in the first and second qudit
Hilbert spaces and are providing the minima of entropies $H_1(u_{10})=S_1$
and $H_2(u_{20})=S_2,$ we obtain the following equality:
\begin{equation}\label{C}
I(u_{10}\otimes u_{20})=S_{1}+S_{2}-H_{12}(u_{10}\otimes u_{20}).
\end{equation}
Since $S_{12}$ is the minimum of $H_{12}(u)$, we have the inequality
$S_{12}\leq H_{12}(u_{10}\otimes u_{20})$, which provides a new inequality
for entropies
\begin{equation}\label{E}
S_{1}+S_{2}\geq H_{12}(u_{10}\otimes u_{20})\geq S_{12}.
\end{equation}
Also we obtain a new inequality for informations $I_q\geq I(u_{10}\otimes
u_{20})$.

For the two-qudit product state with the density matrix
$\rho(1,2)=\rho_1(1)\otimes\rho_2(2)$, we have the equality
$I_q=I(u_{10}\otimes u_{20}).$ Thus, the difference in information
\begin{equation}\label{G}
{\cal D}=(S_{1}+S_{2}-S_{12})-I(u_{10}\otimes u_{20})\geq 0
\end{equation}
is a characteristic of correlations of the qudit subsystems of the
bipartite two-qudit systems. It is an additional characteristic of
correlations in the qudit system, which, in its spirit, is analogous to
discord.

Recently~\cite{Beauty,RitaPS13,JPCS13}, we pointed out that tomograms
$w(m,u)$ and $w(m_1,m_2,u)$ can be interpreted as conditional probability
distributions, i.e.,
$$w(m,u)\equiv w(m\mid u),\qquad w(m_1,m_2,u)\equiv
w(m_1,m_2\mid u).$$ Also for $u=u_1\otimes u_2$, the tomogram
$w(m_1,m_2,u_1,u_2)\equiv w(m_1,m_2\mid u_1,u_2)$.

Thus, all the inequalities discussed can be considered as inequalities for
the entropies and information corresponding to the tomographic conditional
probability distributions.

The properties of Tsallis entropies associated with a joint probability
distribution were discussed in \cite{Rastegin}. We apply these results to
the tomogram $w(m_1,m_2,u)$ of two qudit systems $A$ and $B$. The
tomographic $q$-entropy reads
\begin{equation}\label{A1}
T_q(A,B,u)=\frac{1}{1-q}\left(\sum_{m_1=-j_1}^{j_1}\sum_{m_2=-j_2}^{j_2}w(m_1,m_2,u)^q-1\right).
\end{equation}
In the limit $q\to 1$, this entropy becomes the Shannon tomographic
entropy.

We have the equalities
\begin{equation}\label{A2}
T_q(A,B,u)=T_q(A\mid B,u) +T_q(B,u),
\end{equation}
where $T_q(A\mid B,u)$ is the conditional tomographic $q$-entropy defined
as
\begin{eqnarray}
T_q(A\mid B,u)=\sum_{m_2=-j_2}^{j_2}w(m_2,u)^q~T_q(A\mid m_2,u),\label{A3}\\
T_q(A\mid m_2,u)=\sum_{m_1=-j_1}^{j_1}w(m_1\mid m_2,u)\ln_q
\frac{1}{w(m_1\mid m_2,u)}\,.\label{A4}
\end{eqnarray}
In the last formula (\ref{A4}), the conditional tomographic-probability
distribution $w(m_1\mid m_2,u)$ is defined by the Bayesian formula
$$w(m_1\mid
m_2,u)=\frac{w(m_1,m_2,u)}{\sum_{m_1=-j_1}^{j_1}w(m_1,m_2,u)}.$$
The function $\ln_q(x)$ reads $\ln_q(x)=(x^{1-q}-1)(1-q)^{-1}$, and in the
limit $q\to 1$, $\ln_q(x)=\ln x.$ Also the above relations
(\ref{A1})--(\ref{A4}) become in this limit the relations for Shannon
tomographic entropies.

Using the known inequalities (see, for example, \cite{Rastegin}), we obtain
the inequalities for tomographic entropies
$$T_q(A,u)\leq T_q(A,B,u),\qquad T_q(A\mid B,u)\leq T_q(A,u).$$
A new aspect of these inequalities is that one can consider the minima of
the Tsallis entropy for particular unitary tomograms $u=u_{10}\otimes
u_{20}$ for which
\begin{equation}\label{A5}
T_q(A,u_{10}\otimes
u_{20})=\frac{1}{1-q}\big(\mbox{Tr}\,\rho^q(A)-1\big),\qquad
T_q(A,B,u_{10}\otimes
u_{20})\geq\frac{1}{1-q}\big(\mbox{Tr}\,\rho^q(A,B)-1\big). \end{equation}
In the limit $q\to 1$, $~~T_q(A,u_{10}\otimes u_{20})\to S(A).$

Then we have the inequalities for the von Neumann $S$ and Shannon $H$
entropies as well as the conditional Shannon tomographic entropy $H(A\mid
B,u)$ as follows:
\begin{equation}\label{A6}
S(A)\leq H(A,B,u_{10}\otimes u_{20}),\qquad H(A\mid B,u_{10}\otimes
u_{20})\leq S(A).\end{equation}

\section{Vectors and Matrices with Nonnegative Numbers}
\pst The discussed properties of entropies and their inequalities can be
related to the properties of vectors and matrices. Suppose that one has a
rectangular matrix $P_{jk}$, $j=1,2,\ldots,n$ with nonnegative matrix
elements such that $\sum_{j,k}P_{jk}=1$. This means that one can interpret
the vector $\vec P$ constructed as a column with rows taken as subsequent
pieces of this vector. Among the matrix elements one can have zeros.

An analog of the subadditivity inequality reads
\begin{eqnarray}\label{M1}
&&-\sum_{j=1}^m\left(\sum_{k=1}^nP_{jk}\right)\left(\ln\sum_{k'=1}^nP_{jk'}\right)
-\sum_{k=1}^n\left(\sum_{j=1}^mP_{jk}\right)\left(\ln\sum_{j'=1}^nP_{jj'}\right)\geq
-\sum_{j=1}^m\sum_{k=1}^nP_{jk}\ln P_{jk}.
\end{eqnarray}
One can illustrate this inequality for the probability 4-vector $\vec
P=(a,b,c,d)$. We construct the 2$\times$2 matrix of the form
\begin{eqnarray}\label{M2}
P_{jk}=\left(
          \begin{array}{cc}
            a& b\\
            c& d \\
          \end{array}
        \right).
        \end{eqnarray}
Then inequality (\ref{M1}) reads
\begin{eqnarray}\label{M3}
&&-(a+b)\ln(a+b)-(c+d)\ln(c+d)-(a+c)\ln(a+c)-(b+d)\ln(b+d)\nonumber\\
&&\geq-a\ln a-b\ln b-c\ln c-d\ln d.
\end{eqnarray}
One can use all the permutations of numbers $a,b,c$, and $d$ to obtain
other inequalities, but the right-hand side of (\ref{M3}) is invariant
under the permutation.

Now we suppose that the vector $\vec P$ describes a joint probability
distribution for two coins (subsystems $A$ and $B$)
\begin{equation}\label{M4}
a=w(++),\quad b=w(+-),\quad c=w(-+),\quad d=w(--).
\end{equation}
Then the Shannon entropy
\begin{eqnarray}\label{M5}
H(A,B)=-w(++)\ln w(++)-w(+-)\ln w(+-)-w(-+)\ln w(-+)-w(--)\ln w(--)
\end{eqnarray}
is smaller than the sum of entropies
\begin{eqnarray}\label{M6}
H(A)=-\big(w(++)+w(-+)\big)\ln \big(w(++)+w(+-)\big)
-\big(w(-+)+w(--)\big)\ln \big(w(-+)+w(--)\big)
\end{eqnarray}
and
\begin{eqnarray}\label{M7}
H(B)=-\big(w(++)+w(-+)\big)\ln \big(w(++)+w(+-)\big)
-\big(w(-+)+w(--)\big)\ln \big(w(-+)+w(--)\big),
\end{eqnarray}
i.e.,
\begin{equation}\label{M8}
H(A,B)\leq H(A)+H(B).
\end{equation}
But if we take
\begin{equation}\label{M9}
a=w(++),\quad b=w(--),\quad c=w(+-),\quad d=w(-+),
\end{equation}
the general inequality~(\ref{M3}) provides the inequality for the functions
\begin{eqnarray}\label{M10}
H_1=-\big(w(++)+w(--)\big)\ln \big(w(++)+w(--)\big)
-\big(w(+-)+w(-+)\big)\ln \big(w(+-)+w(-+)\big)
\end{eqnarray}
and
\begin{eqnarray}\label{M11}
H_2=-\big(w(++)+w(+-)\big)\ln \big(w(++)+w(+-)\big)
-\big(w(--)+w(-+)\big)\ln \big(w(--)+w(-+)\big),
\end{eqnarray}
which reads
\begin{equation}\label{M12}
H_1+H_2\geq H(A,B).
\end{equation}

\section{Permutations of Factorized Joint Probability Distributions}
\pst Inequality~(\ref{M12}) is different from inequality~(\ref{M8}). We can
see this difference for the 4-vector, which is the tensor product of two
probability vectors
\begin{eqnarray}\label{M2a}
\vec P=\left(
          \begin{array}{c}
            x\\
            y\\
          \end{array}
        \right)\otimes \left(
          \begin{array}{c}
            \alpha\\
            \beta\\
          \end{array}
        \right).
        \end{eqnarray}
We rewrite (\ref{M4}) as
\begin{equation}\label{M14}
a=x\alpha,\quad b=x\beta,\quad c=y\alpha,\quad d=y\beta.
\end{equation}
In this case, instead of $~H(A,B)=H(A)+H(B)$, which follows from (\ref{M8})
for the joint probability vector, we obtain the inequality, which in terms
of number $\alpha,\beta,x$, and $y$, reads
\begin{eqnarray}\label{M15}
&&-(x\alpha+y\beta)\ln(x\alpha+y\beta)-(x\beta+y\alpha)\ln(x\beta+y\alpha)
-x\ln x-y\ln y\nonumber\\
&&\geq-x\ln x-y\ln y-\alpha\ln\alpha-\beta\ln\beta.
\end{eqnarray}
Since $H(A)=H_1$, the sum $H_1+H_2\geq H(A)+H(B)$ means that
\begin{equation}\label{M16}
H_2\geq H(B).
\end{equation}
The meaning of this new inequality for bipartite systems without
correlations in their subsystems needs to be clarified.

\section{Matrices with Three Indices and the Strong Subadditivity \newline
Condition} \pst Another extension of the strong subadditivity condition can
be formulated in terms of the matrix $P_{jkm}$ with three indices. We
suppose that, for $j=1,2,\ldots n_1$, $k=1,2,\ldots n_2$, and $m=1,2,\ldots
n_3$, all numbers $P_{jkm}$ are nonnegative and satisfy the condition
$\sum_{j=1}^{n_1}\sum_{k=1}^{n_2}\sum_{m=1}^{n_3}P_{jkm}=1$. This means
that initially we have the probability $N$-vector $\vec P$ $(N\leq n_1\cdot
n_2\cdot n_3)$ and apply the labels $jkm$ to all components of the vector.
Then an analog of the strong subadditivity condition reads
\begin{eqnarray} \label{M18}
-\sum_{j=1}^{n_1}\sum_{k=1}^{n_2}\sum_{m=1}^{n_3}P_{jkm}\ln P_{jkm}-
\sum_{k=1}^{n_2}\left(\sum_{j=1}^{n_1}\sum_{m=1}^{n_3}P_{jkm}\right)
\ln \left(\sum_{j'=1}^{n_1}\sum_{m'=1}^{n_2}P_{j'km}\right)\nonumber\\
\leq-\sum_{j=1}^{n_1}\sum_{k=1}^{n_2}\left(\sum_{m'=1}^{n_3}P_{jkm'}\right)
\ln \left(\sum_{m=1}^{n_3}P_{jkm}\right)
-\sum_{k=1}^{n_2}\sum_{m=1}^{n_3}\left(\sum_{j'=1}^{n_1}P_{j'km}\right) \ln
\left(\sum_{j=1}^{n_1}P_{jkm}\right).
\end{eqnarray}

In the case where $P_{jkm}=w(j,k,m)$ is a joint probability distribution
for three random variables (three subsystems $A$, $B$, and $C$ of a
composite system), inequality~(\ref{M18}) is the strong subadditivity
condition.

In the particular case of a system without correlations, this inequality
becomes the equality. Nevertheless, this inequality takes place for any set
$P_{jkm}$ of nonnegative numbers, which can also label the components of
arbitrary probability vectors $\vec P$, even for a single system or for a
system consisting of several subsystems.

Any tomogram for $N$ qudit states $w(m_1,m_2,\ldots,m_N,u)$ with the
density matrix $\rho$ can be considered as the probability
$n$-vector~\cite{SudarPLA}
$$\vec w(u)=\mid uu_0\mid^2\vec\rho,$$
where $\vec\rho=(\rho_1,\ldots,\rho_n)$ is the column vector with
eigenvalues of the density matrix $\rho_k$, $k=1,2,\ldots,n$, and columns
of the unitary matrix $u_0$ are the eigenvectors of $\rho$. If one labels
the vector components of the vector $\vec w$ writing it as a matrix
$P_{jk}$ (if necessary, adding the corresponding number of zero components
to the vector $\vec w$), one obtains the inequality for the unitary matrix.

Analogously, one can label the vector components of the vector $\vec w(u)$
as $P_{jkm}$. In this case, the discussed inequalities are entropic
inequalities for tomograms and also the inequalities for the unitary
matrices.

\section{Conclusions}
\pst We point out our main results presented here.

We formulated some inequalities for sets of nonnegative numbers and
matrices with nonnegative matrix elements.

For qudits, we studied relations between Shannon and $q$-entropies known in
the conventional probability theory. We applied these relations to the
tomographic-probability distributions determining the qudit states. Taking
the minima of the entropies with respect to the local unitary transforms,
we obtained the inequalities containing the von Neumann entropies and their
$q$-generalizations. The new inequalities, such as (\ref{G}), (\ref{A5}),
and (\ref{A6}), can be used to characterize the degree of quantum
correlations.

The obtained new entropic and information inequalities for qudit systems
can be considered as some analogs of the quantum discord properties, which
provide an extra clarification of the properties of quantum correlations.
For continuous variables, the quantum evolution equations for optical
tomograms of quantum systems were obtained in \cite{KorenJRLR} and studied
in \cite{KorenPRA}. The dynamical maps describing the evolution of hybrid
classical--quantum systems were studied in \cite{JPCS13}. The evolution of
tomograms yields the evolution of entropies.

It is worth pointing out that some entropic inequalities for optical
tomograms of photon states were checked experimentally~\cite{BelliniPRA},
in addition to the photon-quadrature uncertainty relations checked in
\cite{PorzioPS}.

We will study the evolution  of continuous variables in the future work.

\section*{Acknowledgments}
\pst This work was partially supported by the Russian Foundation for
Basic Research under Project No.~11-02-00456\_a.

\end{document}